\documentclass[aps,preprint,floats,epsf,epsfig,nofootinbib,letter]{revtex4}
\usepackage{amsfonts}
\usepackage{amsmath}
\usepackage{amssymb}
\usepackage{mathptmx}
\usepackage[dvips]{color}
\usepackage{epsfig}
\usepackage{graphicx}
\begin{document}
\def\be{\begin{eqnarray}}
\def\en{\end{eqnarray}}
\def\la{\langle}
\def\ra{\rangle}
\def\non{\nonumber}
\def\B{{\cal B}}
\def\ov{\overline}
\def\up{\uparrow}
\def\dw{\downarrow}
\def\vp{\varepsilon}
\def\CP{{\it CP}~}
\def\pr{{Phys. Rev.}~}
\def\prl{{ Phys. Rev. Lett.}~}
\def\pl{{ Phys. Lett.}~}
\def\np{{ Nucl. Phys.}~}
\def\zp{{ Z. Phys.}~}
\newcommand{\acp}{\ensuremath{A_{CP}}}

\font\el=cmbx10 scaled \magstep2{\obeylines\hfill August, 2009}

\vskip 1.5 cm

\centerline{\large\bf Resolving $B$-{\it CP} Puzzles in QCD Factorization}
\bigskip
\centerline{\bf Hai-Yang Cheng,$^{1,2}$ and Chun-Khiang Chua$^3$}
\medskip
\centerline{$^1$ Institute of Physics, Academia Sinica}
\centerline{Taipei, Taiwan 115, Republic of China}
\medskip
\medskip
\centerline{$^2$ Physics Department, Brookhaven National
Laboratory} \centerline{Upton, New York 11973}
\medskip
\medskip
\centerline{$^3$ Department of Physics, Chung Yuan Christian University}
\centerline{Chung-Li, Taiwan 320, Republic of China}
\medskip

\bigskip
\centerline{\bf Abstract}
\bigskip
\small
Within the framework of QCD factorization (QCDF), power
corrections due to penguin annihilation can account for the
observed  rates of penguin-dominated two-body decays of $B$ mesons
and direct \CP asymmetries $\acp(K^-\pi^+)$, $\acp(K^{*-}\pi^+)$,
$\acp(K^-\rho^0)$ and $\acp(\pi^+\pi^-)$. However, the predicted
direct {\it CP}-violating effects in QCDF for $B^-\to K^-\pi^0,K^-\eta,\pi^-\eta$ and $\bar B^0\to\pi^0\pi^0$ are wrong in signs
when confronted with experiment. We show that subleading $1/m_b$
power corrections to the color-suppressed tree amplitude due to
spectator scattering or final-state interactions will yield
correct signs for  aforementioned \CP asymmetries  and accommodate
the observed $\pi^0\pi^0$ and $\rho^0\pi^0$ rates simultaneously.
Implications are discussed.

\newpage
\par {\bf 1.}
In the heavy quark limit, hadronic matrix elements can be
expressed in terms of certain nonperturbative input quantities
such as light cone distribution amplitudes and transition form
factors. Consequently, the decay amplitudes of charmless two-body
decays of $B$ mesons can be described  in terms of decay constants
and form factors. However, the predicted rates for
penguin-dominated $B\to PP,VP,VV$ decays ($P$ and $V$ denoting
pseudoscalar and vector mesons, respectively) are  systematically
below the measurements (see the second column of Table
\ref{tab:PP}; for a review, see \cite{ChengSmith}).\footnote{We
have included chirally enhanced but power suppressed penguin
contributions. Numerically, they are of order $1/m_b^0$.}
Moreover, the calculated direct \CP asymmetries for $\bar B^0\to
K^-\pi^+,K^{*-}\pi^+$, $B^-\to K^-\rho^0$ and $\bar
B^0\to\pi^+\pi^-$ are wrong in signs when confronted with
experiment as shown in the same Table. This implies the necessity
of taking into account $1/m_b$ power correction effects. In the
QCD factorization (QCDF) approach \cite{BBNS}, power corrections
often involve endpoint divergences. For example,  the hard
spectator scattering diagram at twist-3 order is power suppressed
and posses soft and collinear divergences arising from the soft
spectator quark and the $1/m_b$ annihilation amplitude has
endpoint divergences even at twist-2 level. Since the treatment of
endpoint divergences is model dependent, subleading power
corrections generally can be studied only in a phenomenological
way. While the endpoint divergence is regulated in the pQCD
approach by introducing the parton's transverse momentum
\cite{pQCD},  it is parameterized in QCD factorization as
 \be
 X_A\equiv \int^1_0{dy\over y}={\rm ln}{m_B\over
 \Lambda_h}(1+\rho_A e^{i\phi_A}),
 \end{eqnarray}
for penguin annihilation contributions
with $\Lambda_h$ being a typical scale of order
500 MeV.

In the so-called ``S4" scenario of QCDF \cite{BN} with some
appropriate choice of the parameters $\rho_A$ and $\phi_A$, the
above-mentioned discrepancies are resolved in the presence of
power corrections due to the penguin annihilation topology.
However, a scrutiny of the QCDF predictions reveals more puzzles
in the regard of direct \CP violation. When power corrections due
to penguin annihilation are turned on, the signs of $\acp$ in
$B^-\to K^-\pi^0,~K^-\eta,~\pi^-\eta$ and $\bar B^0\to\pi^0\pi^0$
will also get flipped in such a way that they disagree with
experiment (see the third column of Table \ref{tab:PP}). The
so-called $K\pi$ {\it CP}-puzzle is related to the difference of
\CP asymmetries of $B^-\to K^-\pi^0$ and $\bar B^0\to K^-\pi^+$.
This can be illustrated by considering the decay amplitudes of
$\bar B\to \bar K\pi$ in terms of topological diagrams
 \begin{eqnarray} \label{eq:ampBKpi}
 A(\bar B^0\to K^-\pi^+) &=& P'+T'+{2\over 3}P'^c_{\rm EW}+P'_A, \nonumber \\
 A(\bar B^0\to \bar K^0\pi^0) &=& -{1\over \sqrt{2}}(P'-C'-P'_{\rm EW}-{1\over 3}P'^c_{\rm EW}+P'_A), \\
 A(B^-\to \bar K^0\pi^-) &=& P'-{1\over 3}P'^c_{\rm EW}+A'+P'_A,  \nonumber \\
 A(B^-\to K^-\pi^0) &=& {1\over\sqrt{2}}(P'+T'+C'+P'_{\rm
 EW}+{2\over 3}P'^c_{\rm EW}+A'+P'_A), \non
 \end{eqnarray}
where $T$, $C$, $E$, $A$, $P_{\rm EW}$ and $P^c_{\rm EW}$ are
color-allowed tree, color-suppressed tree, $W$-exchange,
$W$-annihilation, color-allowed and color-suppressed electroweak
penguin amplitudes, respectively, and $P_A$ is the penguin-induced
weak annihilation amplitude. We use unprimed and primed symbols to
denote $\Delta S=0$ and $|\Delta S|=1$ transitions. We notice that
if $C'$, $P'_{\rm EW}$ and $A'$ are negligible compared with $T'$,
it is clear from Eq. (\ref{eq:ampBKpi}) that the decay amplitudes
of $K^-\pi^0$ and $K^-\pi^+$ will be the same apart from a trivial
factor of $1/\sqrt{2}$. Hence, one will expect that
$\acp(K^-\pi^0)\approx \acp(K^-\pi^+)$, while they differ by
5.3$\sigma$ experimentally, $\Delta
A_{K\pi}\equiv\acp(K^-\pi^0)-\acp(K^-\pi^+)=0.148\pm0.028$
\cite{HFAG}. We also notice that the decay $B^-\to K^-\eta$ has a
world average $-0.37\pm0.09$ for $\acp(K^-\eta)$
\cite{HFAG,PDG,BaBar:Beta} different from zero by 4.1 standard
deviations.

Since in the heavy quark limit, \CP asymmetries of the
$K^-\pi^0,~K^-\eta,~\pi^-\eta,~\pi^0\pi^0$ modes have the correct signs when
compared with experiment, the $B$-{\it CP} puzzles mentioned here
are relevant to QCDF and may not occur in other approaches such as
pQCD. In this work, we shall show that soft power corrections to
the color-suppressed tree amplitude will bring the signs of $\acp$
back to the right track. As a bonus, the rates of $\bar
B^0\to\pi^0\pi^0,~\rho^0\pi^0$ can be accommodated.

\vskip 0.4cm {\bf 2.}~The aforementioned direct \CP puzzles
indicate that it is necessary to consider subleading power
corrections other than penguin annihilation. For example, the
large power corrections due to $P'_A$ cannot explain the $\Delta
A_{K\pi}$ puzzle as they contribute equally to both $B^-\to
K^-\pi^0$ and $\bar B^0\to K^-\pi^+$. The additional power
correction should have little impact on the decay rates of
penguin-dominated decays but will manifest in the measurement of
direct \CP asymmetries. Note that all the "problematic" modes
receive a contribution from $c^{(')}=C^{(')}+P_{\rm EW}^{(')}$.
Since $A(B^-\to K^-\pi^0)\propto t'+c'+p'$ and $A(\bar B^0\to
K^-\pi^+)\propto t'+p'$ with $t'=T'+P'^c_{\rm EW}$ and
$p'=P'-{1\over 3}P'^c_{\rm EW}+P'_A$, we can consider this puzzle
resolved, provided that $c'/t'$ is of order $1.3\sim 1.4$ with a
large negative phase (naively $|c'/t'|\sim 0.9$). There are several possibilities for a large $c'$: either a large color suppressed
$C'$ or a large electroweak penguin $P'_{\rm EW}$ or a combination of them. Various
scenarios for accommodating large $C'$
\cite{Li05,Kim,Rosner,Chua,Ciuchini,Kagan,Li09,Baek09} or $P'_{\rm
EW}$ \cite{LargeEWP,Hou} have been proposed.
%
%
To get a large $C'$, one can appeal to spectator scattering or
final-state rescattering (see discussions below). However, the
general consensus for a large $P'_{\rm EW}$ is that one needs New
Physics beyond the Standard Model. In principle, one cannot tell
the difference of these two possibilities in penguin-dominated
decays as it is always the combination $c'=C'+P'_{\rm EW}$ that
enters into the decay amplitude except for the decays involving
$\eta$ and/or $\eta'$ in the final state where
%
%
both $c'$ and $P'_{\rm EW}$ present in the
amplitudes~\cite{Chiang}.
Nevertheless, the two scenarios can lead to very distinct
predictions for tree-dominated decays where $P_{\rm EW}\ll C$ as
the electroweak penguin amplitude here does not get a CKM enhancement. The
decay rates of $\bar B^0\to \pi^0\pi^0,\rho^0\pi^0$ will be
substantially enhanced for a large $C$ but remain intact for a
large $P'_{\rm EW}$. Since $P_{\rm EW}\ll C$ in tree-dominated
channels, \CP puzzles with $\pi^-\eta$ and $\pi^0\pi^0$ cannot be
resolved with a large $P'_{\rm EW}$. Therefore, it is most likely
that the  color-suppressed tree amplitude is large and complex. Motivated by the above observation, in this work we shall  consider the possibility of a large complex
$a_2$, the parameter for describing the color-suppressed tree
topology, and parameterize power corrections to $a_2$, as
\footnote{We use NLO results for $a_2$ in Eq. (\ref{eq:a2}) as a
benchmark to define power corrections. The NNLO calculations of
spectator-scattering tree amplitudes and vertex corrections at
order $\alpha_s^2$ have been carried out in \cite{Beneke} and
\cite{Bell}, respectively. While NNLO corrections can in principle
push the magnitude of $a_2(\pi\pi)$ up to the order of 0.50 by
lowering the value of the $B$-meson parameter $\lambda_B$, the
strong phase of $a_2$ relative to $a_1$ cannot be larger than
$15^\circ$ \cite{BenekeFPCP}.}
 \be \label{eq:a2}
 a_2 \to a_2^{\rm NLO}(1+\rho_C e^{i\phi_C}),
 \en
with the unknown parameters $\rho_C$ and $\phi_C$ to be inferred
from experiment.

The reader is referred to \cite{CCSnew} for details. We shall
first consider soft corrections to weak annihilation dictated by
the parameters $\rho_A$ and $\phi_A$. A fit to the data of
two-body hadronic decays of $B^0$ and $B^-$ mesons within QCDF
yields the values
 \be \label{eq:rhoA}
\rho_A^0 &\approx& 1.10,\quad 1.07,\quad 0.87, \non \\
\phi_A^0 &\approx& -50^\circ,~-70^\circ,~-30^\circ,
 \en
for $B\to PP,PV,VP$ respectively, where the superscript ``0" of
$\rho_A$ and $\phi_A$ indicates that they are the default values we
shall use in this work. Basically, this is very similar to the
``scenario S4" presented in \cite{BN}. For the annihilation
diagram we use the convention that $M_1$ ($M_2$) contains an
antiquark (a quark) from the weak vertex.  Since the penguin
annihilation effects are different for $M_1=P$ and $M_1=V$, the
parameters $\rho_A$ and $\phi_A$ are thus different for $B\to PV$
and $B\to VP$.

Branching fractions and direct \CP asymmetries for  some selective $B\to PP$
decays are shown in Table \ref{tab:PP}. The theoretical errors
correspond to the uncertainties due to variation of (i) the
Gegenbauer moments, the  decay constants, (ii) the heavy-to-light
form factors and the strange quark mass, and (iii) the wave
function of the $B$ meson characterized by the parameter
$\lambda_B$, the power corrections due to weak annihilation and
hard spectator interactions described by the parameters
$\rho_{A,H}$, $\phi_{A,H}$, respectively. To obtain the errors
shown in Table \ref{tab:PP}, we first scan randomly the points in
the allowed ranges of the above nine parameters (specifically, the
ranges $\rho_A^0-0.1\leq \rho_A\leq \rho_A^0+0.1$,
$\phi_A^0-20^\circ\leq \phi_A\leq \phi_A^0+20^\circ$, $0\leq
\rho_H\leq 1$ and $0\leq \phi_{H}\leq 2\pi$ are used in this work)
and then add errors in quadrature. More specifically, the second
error in the table is referred  to the uncertainties caused by the
variation of  $\rho_{A,H}$ and $\phi_{A,H}$, where all other
uncertainties are lumped into the first error.  Power corrections
beyond the heavy quark limit generally give the major theoretical
uncertainties.

 \begin{table}[t]
 \caption{$CP$-averaged branching fractions (in units of $10^{-6}$) and direct \CP
asymmetries (in \%) of some selective $B\to PP$ decays obtained in
QCD factorization for three distinct cases: (i) without any power
corrections, (ii) with power corrections from penguin
annihilation, and (iii) with power corrections to both penguin
annihilation and color-suppressed tree amplitudes. The parameters
$\rho_A$ and $\phi_A$ are taken from Eq. (\ref{eq:rhoA}),
 $\rho_C=1.3$ and $\phi_C=-70^\circ$. Sources of theoretical uncertainties are discussed in the text.
 } \label{tab:PP}
\begin{ruledtabular}
 \begin{tabular}{l  c c c c c}
 {Modes}
   &   {\rm W/o}~$\rho_{A,C},\phi_{A,C}$ &  {\rm With}~$\rho_A,\phi_A$ &  {\rm With}~$\rho_{A,C},\phi_{A,C}$ & Expt. \cite{HFAG} \\  \hline
   $\B({\overline{B}}^{0}\to K^{-}\pi^+)$
        & $13.1^{+5.8+0.7}_{-3.5-0.7}$
                                                                &
                                                                $19.3^{+7.9+8.2}_{-4.8-6.2}$
                                                                & $19.3^{+7.9+8.2}_{-4.8-6.2}$  & $19.4\pm0.6$
                                                                \\
   $\B({\overline{B}}^{0} \to \bar K^{0}{\pi}^{0})$
     & $5.5^{+2.8+0.3}_{-1.7-0.3}$
                                                                 &
                                                                 $8.4^{+3.8+3.8}_{-2.3-2.9}$
                                                                 & $8.6^{+3.8+3.8}_{-2.2-2.9}$  & $9.8\pm0.6$
                                                                 \\
  $\B(B^-\to \bar K^0\pi^-)$ & $14.9^{+6.9+0.9}_{-4.5-1.0}$ & $21.7^{+9.2+9.0}_{-6.0-6.9}$ & $21.7^{+9.2+9.0}_{-6.0-6.9}$ & $23.1\pm1.0$ \\
  $\B(B^-\to K^-\pi^0)$ & $9.1^{+3.6+0.5}_{-2.3-0.5}$ & $12.6^{+4.7+4.8}_{-3.0-3.7}$ &  $12.5^{+4.7+4.9}_{-3.0-3.8}$ & $12.9\pm0.6$ \\
  $\B(B^-\to K^-\eta)$ & $1.6^{+1.1+0.3}_{-0.7-0.4}$ & $2.4^{+1.8+1.3}_{-1.1-1.0}$ & $2.4^{+1.8+1.3}_{-1.1-1.0}$ & $2.3\pm0.3$ \footnotemark[1] \\
  $\B({\overline{B}}^{0}\to \pi^{+}\pi^-)$ & $6.2^{+0.4+0.2}_{-0.6-0.4}$ & $7.0^{+0.4+0.7}_{-0.7-0.7}$ & $7.0^{+0.4+0.7}_{-0.7-0.7}$ & $5.16\pm0.22$ \\
  $\B({\overline{B}}^{0}\to \pi^{0}\pi^0)$ & $0.42^{+0.29+0.18}_{-0.11-0.08}$ & $0.52^{+0.26+0.21}_{-0.10-0.10}$ & $1.1^{+1.0+0.7}_{-0.4-0.3}$ & $1.55\pm0.19$ \footnotemark[2] \\
  $\B({{B}}^{-}\to \pi^{-}\pi^0)$ & $4.9^{+0.9+0.6}_{-0.5-0.3}$ & $4.9^{+0.9+0.6}_{-0.5-0.3}$ & $5.9^{+2.2+1.4}_{-1.1-1.1}$ & $5.59^{+0.41}_{-0.40}$ \\
  $\B(B^-\to \pi^-\eta)$ & $4.4^{+0.6+0.4}_{-0.3-0.2}$ & $4.5^{+0.6+0.5}_{-0.3-0.3}$ &  $5.0^{+1.2+0.9}_{-0.6-0.7}$ & $4.1\pm0.3$ \footnotemark[1] \\
  \hline
   $\acp({\overline{B}}^{0}\to K^{-}\pi^+)$
        & $4.0^{+0.6+1.1}_{-0.7-1.1}$
                                                                &
                                                                $-7.4^{+1.7+4.3}_{-1.5-4.8}$
                                                                & $-7.4^{+1.7+4.3}_{-1.5-4.8}$  & $-9.8^{+1.2}_{-1.1}$
                                                                \\
   $\acp({\overline{B}}^{0} \to \bar K^{0}{\pi}^{0})$
     & $-4.0^{+1.2+3.5}_{-1.8-3.0}$
                                                                 &
                                                                 $0.75^{+1.88+2.56}_{-0.94-3.32}$
                                                                 & $-10.6^{+2.7+5.6}_{-3.8-4.3}$  & $-1\pm10$ 
                                                                 \\
  $\acp(B^-\to \bar K^0\pi^-)$ & $0.72^{+0.06+0.05}_{-0.05-0.05}$ & $0.28^{+0.03+0.09}_{-0.03-0.10}$ & $0.28^{+0.03+0.09}_{-0.03-0.10}$ & $0.9\pm2.5$ \\
  $\acp(B^-\to K^-\pi^0)$ & $7.3^{+1.6+2.3}_{-1.2-2.7}$ & $-5.5^{+1.3+4.9}_{-1.8-4.6}$ & $4.9^{+3.9+4.4}_{-2.1-5.4}$ & $5.0\pm2.5$ \\
  $\acp(B^-\to K^-\eta)$ & $-22.1^{+7.7+14.0}_{-16.7-~7.3}$ & $12.7^{+7.7+13.4}_{-5.0-15.0}$ & $-11.0^{+~8.4+14.9}_{-21.6-10.1}$ & $-37\pm9$ \footnotemark[1] \\
  $\acp({\overline{B}}^{0}\to \pi^{+}\pi^-)$ & $-6.2^{+0.4+2.0}_{-0.5-1.8}$ & $17.0^{+1.3+4.3}_{-1.2-8.7}$ & $17.0^{+1.3+4.3}_{-1.2-8.7}$ & $38\pm6$ \\
  $\acp({\overline{B}}^{0}\to \pi^{0}\pi^0)$ & $33.4^{+~6.8+34.8}_{-10.6-37.7}$ & $-26.9^{+8.4+48.5}_{-6.0-37.5}$ & $57.2^{+14.8+30.3}_{-20.8-34.6}$ & $43^{+25}_{-24}$ \\
  $\acp({{B}}^{-}\to \pi^{-}\pi^0)$ & $-0.06^{+0.01+0.01}_{-0.01-0.02}$ & $-0.06^{+0.01+0.01}_{-0.01-0.02}$ &
  $-0.11^{+0.01+0.06}_{-0.01-0.03}$ & $6\pm5$ \\
  $\acp(B^-\to \pi^-\eta)$ & $-11.4^{+1.1+2.3}_{-1.0-2.7}$ & $11.4^{+0.9+4.5}_{-0.9-9.1}$ & $-5.0^{+2.4+~8.4}_{-3.4-10.3}$ & $-13\pm7$ \footnotemark[1]\\
 \end{tabular}
\footnotetext[1]{We have taken into account the new measurement of $B^-\to (K^-,\pi^-)\eta$ \cite{BaBar:Beta} to update the average.}
\footnotetext[2]{This is the average of $1.83\pm0.21\pm0.13$ by BaBar \cite{BaBarpi0pi0} and $1.1\pm0.3\pm0.1$ by Belle \cite{Bellepi0pi0}. If an $S$ factor is included, the average will become $1.55\pm0.35$\,.}
 \end{ruledtabular}
 \end{table}

For $\rho_C\approx 1.3 $ and $\phi_C\approx -70^\circ$, we find
that all the \CP puzzles in $B\to PP$ decays are resolved as shown
in fourth column of Table \ref{tab:PP}. The corresponding $a_2$'s
are
 \be \label{eq:a2PP}
 a_2(\pi \pi)\approx 0.60\,e^{-i55^\circ},
 &\quad&
 a_2(K\pi)\approx 0.51\,e^{-i58^\circ}.
 \en
They are consistent with the phenomenological determination of
$C^{(')}/T^{(')}\sim a_2/a_1$ from a global fit to the available
data \cite{Chiang}. Due to the interference between the penguin
and the enhanced color-suppressed amplitudes with a sizable strong
phase, it is clear from Table \ref{tab:PP} that theoretical
predictions for $\acp$ now agree with experiment in sign even for
those modes with the measured $\acp$ less than 3$\sigma$ in significance. As first
emphasized by Lunghi and Soni \cite{Lunghi}, in the QCDF analysis
of the quantity $\Delta A_{K\pi}$, although the theoretical
uncertainties due to power corrections from penguin annihilation
are large for individual asymmetries $\acp(K^-\pi^0)$ and
$\acp(K^-\pi^+)$, they essentially cancel out in their difference,
rendering the theoretical prediction more reliable. We find
$\Delta A_{K\pi}=(12.3^{+2.2+2.1}_{-0.9-4.7})\%$, while it is only
$(1.9^{+0.5+1.6}_{-0.4-1.0})\%$ in the absence of power
corrections to  the topological amplitude ``$C$" or $a_2$.

For the direct \CP asymmetry of $\bar B^0\to \bar K^0\pi^0$, we
predict $\acp(\bar K^0\pi^0)=(-10.6^{+2.7+5.5}_{-3.7-4.3})\%$.
Experimentally, the current world average $-0.01\pm0.10$ is
consistent with no \CP violation because the BaBar and Belle
measurements, $-0.13\pm0.13\pm0.03$ \cite{BaBarK0pi0} and
$0.14\pm0.13\pm0.06$ \cite{BelleK0pi0} respectively, are opposite
in sign.  Nevertheless, there exist several model-independent
determinations of this asymmetry: one is the SU(3) relation
    $\Delta \Gamma(\pi^0\pi^0)=-\Delta \Gamma(\bar K^0\pi^0)$ \cite{Deshpande},
and the other is the approximate sum rule for \CP rate asymmetries \cite{AS98}
\begin{eqnarray} \label{eq:SR}
\Delta\Gamma(K^-\pi^+)+\Delta \Gamma(\bar K^0\pi^-)\approx 2[\Delta \Gamma(K^-\pi^0)+\Delta \Gamma(\bar K^0\pi^0)],
\end{eqnarray}
based on isospin symmetry, where $\Delta \Gamma(K\pi)\equiv
\Gamma(\bar B\to\bar K\bar\pi)-\Gamma(B\to K\pi)$. This sum rule
allows us to extract $\acp(\bar K^0\pi^0)$ in terms of the other
three asymmetries in $K^-\pi^+,K^-\pi^0,\bar K^0\pi^-$ modes that
have been measured. From the current data of branching fractions
and \CP asymmetries, the above SU(3) relation and {\it
CP}-asymmetry sum rule lead to $\acp(\bar
K^0\pi^0)=-0.073^{+0.042}_{-0.041}$ and $\acp(\bar
K^0\pi^0)=-0.15\pm 0.04$, respectively. An analysis based on the
topological quark diagrams also yields a similar result $-0.08\sim
-0.12$ \cite{Chiang09}. All these indicate that the direct \CP
violation  $\acp(\bar K^0\pi^0)$ should be negative and has a
magnitude  of order 0.10\,. As for the mixing-induced asymmetry
$S_{\pi^0K_S}$, it is found to be enhanced from 0.76 to
$0.79^{+0.06+0.04}_{-0.04-0.04}$ when $\rho_C$ and $\phi_C$ are
turned on, while experimentally it is $0.57\pm0.17$ \cite{HFAG}.
The discrepancy between theory and experiment for $S_{\pi^0K_S}$
is one of possible hints of New Physics \cite{CCSsin2beta}. Our
result for $S_{\pi^0 K_S}$ is consistent with
\cite{Chua,Ciuchini,Kagan} where soft corrections to $a_2$ were
considered, but not with \cite{Li09} where $S_{\pi^0 K_S}\sim
0.63$ was obtained. A correlation between $S_{\pi^0 K_S}$ and
$\acp(\pi^0K_S)$ has been investigated recently in
\cite{Fleischer}. For the mixing-induced asymmetry in $B\to
\pi^+\pi^-$, we find
$S_{\pi^+\pi^-}=-0.69^{+0.08+0.19}_{-0.10-0.09}$, in accordance
with the world average of $-0.65\pm0.07$ \cite{HFAG}.

From Table \ref{tab:PP} we see that power corrections to the
color-suppressed tree amplitude have almost no impact on the decay
rates of penguin-dominated decays, but will enhance the
color-suppressed tree dominated decay $B\to\pi^0\pi^0$
substantially owing to the enhancement of $|a_2|\sim {\cal
O}(0.6)$.  Notice that the central values of the branching
fractions of this mode measured by BaBar \cite{BaBarpi0pi0} and
Belle \cite{Bellepi0pi0} are somewhat different as noticed in
Table \ref{tab:PP}. It is generally believed that direct \CP
violation of $B^-\to\pi^-\pi^0$ is very small. This is because the
isospin of the $\pi^-\pi^0$ state is $I=2$ and hence it does not
receive QCD penguin contributions and receives only the loop
contributions from electroweak penguins. Since this decay is tree
dominated, SM predicts an almost null \CP asymmetry, of order
$10^{-3}\sim 10^{-4}$. What will happen if $a_2$ has a large
magnitude and strong phase ? We find that soft corrections to the
color-suppressed tree amplitude will enhance $\acp(\pi^-\pi^0)$
substantially to the level of 2\%. Similar conclusions were also
obtained by the analysis based on the diagrammatic approach
\cite{Chiang}. However, one must be very cautious about this. The
point is that power corrections will affect not only $a_2$, but
also other parameters $a_i$ with $i\neq 2$. Since the isospin of
$\pi^-\pi^0$ is $I=2$,  soft corrections to $a_2$ and $a_i$  must
be conspired in such a way that $\pi^-\pi^0$  is always an $I=2$
state. As explained below, there are two possible sources of power
corrections to $a_2$: spectator scattering and final-state
interactions.  For final-state rescattering, it is found in
\cite{CCSfsi} that effects of FSIs on $\acp(\pi^-\pi^0)$ are
small, consistent with the requirement followed from the CPT
theorem. In the specific residual scattering model considered by
one of us (CKC) \cite{Chua}, $\pi^-\pi^0$ can only rescatter into
itself, and as a consequence, direct \CP violation will not
receive any contribution from final-state interactions.  Likewise,
if large $\rho_H$ and $\phi_H$ are turned on to mimic Eq.
(\ref{eq:a2PP}), we find $\acp(\pi^-\pi^0)$ is at most of order
$10^{-3}$. This is because spectator scattering contributes to
not only $a_{2}$ but also $a_1$ and the electroweak penguin
parameters $a_{7-10}$. Therefore,  a measurement of direct \CP
violation in $B^-\to \pi^-\pi^0$ still provides a nice test of the
Standard Model and New Physics.

In order to explain \CP violation in the decay $B^-\to K^-\eta$,
we shall elaborate it in more detail. Its decay amplitude is given
by \cite{BN}
 \be
\sqrt{2}A(B^-\to K^-\eta) &=& A_{\bar K\eta_q}\left[\delta_{pu}\alpha_2+2\alpha_3^p+{1\over 2}\alpha_{\rm 3,EW}^p \right] \non \\
&+&\sqrt{2}A_{\bar K\eta_s}\left[\delta_{pu}\beta_2+\alpha_3^p+\alpha_4^p-{1\over 2}\alpha_{\rm 3,EW}^p-{1\over 2}\alpha_{\rm 4,EW}^p+\beta_3^p+\beta_{\rm 3,EW}^p\right]  \\
&+&\sqrt{2}A_{\bar K\eta_c}\left[\delta_{pc}\alpha_2+\alpha_3^p\right]
 +A_{\eta_q \bar K}\left[\delta_{pu}(\alpha_1+\beta_2)+\alpha_4^p+\beta_3^p+\beta_{\rm 3,EW}^p\right], \non
 \en
where the flavor states of the $\eta$ meson,  $q\bar q\equiv
(u\bar u+d\bar d)/\sqrt{2}$, $s\bar s$ and $c\bar c$ are labelled
by the $\eta_q$, $\eta_s$ and $\eta_{c}^0$, respectively. The
reader is referred to \cite{BN} for other notations. The physical
states $\eta,~\eta'$ and $\eta_c$ can be expressed in terms of
flavor states $\eta_q$, $\eta_s$ and $\eta_{c}^0$. Since the two
penguin processes $b\to ss\bar s$ and $b\to sq\bar q$ contribute
destructively to $B\to K\eta$, the penguin amplitude is comparable
in magnitude to the tree amplitude induced from $b\to us\bar u$,
contrary to the decay $B\to K\eta'$ which is dominated by large
penguin amplitudes. Consequently, a sizable direct \CP asymmetry
is expected in $B^-\to K^-\eta$ but not in $K^-\eta'$ \cite{BSS}.

Quantities relevant to the calculation are the decay constants
$f_\eta^{q}$, $f_\eta^{s}$ and $f_\eta^c$ defined by $\la 0|\bar
q\gamma_\mu\gamma_5q|\eta\ra=if_\eta^q/\sqrt{2} q_\mu$, $\la
0|\bar s\gamma_\mu\gamma_5s|\eta\ra=if_\eta^s q_\mu$ and $\la
0|\bar c\gamma_\mu\gamma_5c|\eta\ra=if_\eta^c q_\mu$,
respectively. A straightforward perturbative calculation gives
\cite{fetac}
 \be f_\eta^c=-{m_\eta^2\over 12
m_c^2}\,{f_\eta^q\over\sqrt{2}}.
 \en
For the decay constants $f_\eta^q$ and $f_\eta^s$, we shall use
the values $f_\eta^q=107$ MeV and $f_\eta^s=-112$ MeV obtained in
\cite{FKS} with the convention of $f_\pi=132$ MeV. Although the
decay constant $f_\eta^c\approx -2$ MeV is much smaller than
$f_\eta^{q,s}$, its effect is CKM enhanced by
$V_{cb}V_{cs}^*/(V_{ub}V_{us}^*)$. In the absence of  power
corrections to $a_2$, $\acp(K^-\eta)$ is found to be $0.127$ (see
Table \ref{tab:PP}). When $\rho_C$ and $\phi_C$ are turned on,
$\acp(K^-\eta)$ will be reduced to 0.004 if there is no  intrinsic
charm content in the $\eta$. When the effect of $f_\eta^c$ is
taken into account, $\acp(K^-\eta)$ finally reaches at the level
of $-11\%$ and has a sign in agreement with experiment. Hence, \CP
violation in $B^-\to K^-\eta$ is the place where the charm content
of the $\eta$ plays a role.

We add a remark here that the pQCD prediction for $\acp(K^-\eta)$
is very sensitive to $m_{qq}$, the mass of the $\eta_q$, which is
generally taken to be of order $m_\pi$. It was found in
\cite{ChenKeta} that for $m_{qq}=0.14$, 0.18 and 0.22 GeV,
$\acp(K^-\eta)$ becomes 0.0562, 0.0588 and $-0.3064$,
respectively. There are two issues here: (i) Is it natural to have
a large value of $m_{qq}$ ? and (ii) The fact that $\acp(K^-\eta)$
is so sensitive to $m_{qq}$ implies that the pQCD prediction is
not stable. Within the framework of pQCD, the authors of
\cite{Xiao} rely on the NLO corrections to get a negative \CP
asymmetry and to avoid the aforementioned issues. At the lowest
order, pQCD predicts $\acp(K^-\eta)\approx 9.3\%$. Then NLO
corrections will change the sign and give rise to
$\acp(K^-\eta)=(-11.7^{+~8.4}_{-11.4})\%$ \cite{Xiao}.

As for the decay $B^-\to\pi^-\eta$, it is interesting to see that
penguin annihilation will flip the sign of $\acp(\pi^-\eta)$ into
a wrong one without affecting its magnitude (see Table
\ref{tab:VP}). Again, soft corrections to $a_2$ will bring the \CP
asymmetry back to the right track. Contrary to the previous case,
the charm content of the $\eta$ here does not play a
role as it does not get a CKM enhancement relative to the
non-charm content of the $\eta$. Our result of $\acp(\pi^-\eta)=
-0.05^{+0.09}_{-0.11}$ is consistent with the measurement of
$-0.13\pm0.07$. For comparison, the pQCD approach predicts
$-0.37^{+0.09}_{-0.07}$ \cite{Xiaopieta} and SCET gives two
solutions \cite{Zupan},  $0.05\pm0.29$ and $0.37\pm0.29$ with signs opposite to the data.

\vskip 0.4cm
{\bf 3.}~What is the origin of power corrections to
$a_2$ ? There are two possible sources: spectator scattering and
final-state interactions. The flavor operators $a_i^{p}$ are
basically the Wilson coefficients in conjunction with
short-distance nonfactorizable corrections such as vertex
corrections, penguin contractions and hard spectator interactions.
In general, they have the expression \cite{BBNS,BN}
 \be \label{eq:ai}
  a_i^{p}(M_1M_2) &=&
 \left(c_i+{c_{i\pm1}\over N_c}\right)N_i(M_2)
   + {c_{i\pm1}\over N_c}\,{C_F\alpha_s\over
 4\pi}\Big[V_i(M_2)+{4\pi^2\over N_c}H_i(M_1M_2)\Big]+P_i^{p}(M_2),
 \en
where $i=1,\cdots,10$,  the upper (lower) signs apply when $i$ is
odd (even), $c_i$ are the Wilson coefficients,
$C_F=(N_c^2-1)/(2N_c)$ with $N_c=3$, $N_i(M_2)=0$ for $i=6,8$ and equals to 1 otherwise, $M_2$ is the emitted meson
and $M_1$ shares the same spectator quark with the $B$ meson. The
quantities $V_i(M_2)$ account for vertex corrections,
$H_i(M_1M_2)$ for hard spectator interactions with a hard gluon
exchange between the emitted meson and the spectator quark of the
$B$ meson and $P_i(M_2)$ for penguin contractions. A typical
hard spectator term $H_i(M_1 M_2)$ has the expressions \cite{BBNS,BN}:
\begin{eqnarray}\label{eq:hardspec}
  H_i(M_1 M_2)= {if_B f_{M_1} f_{M_2} \over X^{(\overline{B} M_1,
  M_2)}}\,{m_B\over\lambda_B} \int^1_0 d x d y \,
 \Bigg( \frac{\Phi_{M_1}(x) \Phi_{M_2}(y)}{\bar x\bar y} + r_\chi^{M_1}
  \frac{\Phi_{m_1} (x) \Phi_{M_2}(y)}{\bar x y}\Bigg),
 \hspace{0.5cm}
 \end{eqnarray}
for $i=1-4,9,10$, where $X^{(\overline{B} M_1, M_2)}$ is the
factorizable amplitude for $\ov B\to M_1M_2$, $\bar x=1-x$,
$\lambda_B$ is the fist inverse moment of the $B$ meson light-cone
wave function and \be \label{eq:rchi}
 r_\chi^P(\mu)={2m_P^2\over m_b(\mu)(m_2+m_1)(\mu)},  \qquad r_\chi^V(\mu) = \frac{2m_V}{m_b(\mu)}\,\frac{f_V^\perp(\mu)}{f_V} \,.
\en
Power corrections from the twist-3 amplitude $\Phi_m$ are divergent and can be parameterized as
 \be
 X_H\equiv \int^1_0{dy\over y}={\rm ln}{m_B\over
 \Lambda_h}(1+\rho_H e^{i\phi_H}),
 \end{eqnarray}
Since $c_1\sim {\cal O}(1)$ and $c_9\sim {\cal O}(-1.3)$ in units
of $\alpha_{em}$, it turns out that spectator scattering
contributions to $a_i$ are usually small except for $a_2$ and
$a_{10}$ which are essentially governed by hard spectator
interactions \cite{BBNS2}. The value $a_2(K\pi)\approx
0.51\,e^{-i58^\circ}$ corresponds to $\rho_H\approx 4.9$ and
$\phi_H\approx -77^\circ$. \footnote{As pointed out in
\cite{BenekeFPCP,Bell09}, a smaller value of $\lambda_B$ of order
200 MeV can enhance the hard spectator interaction [see Eq.
(\ref{eq:hardspec})] and hence $a_2$ substantially. However,
the recent BaBar data on $B\to\gamma \ell\bar\nu$
\cite{BaBar:gammalnu} seems to imply a larger $\lambda_B$ ($>
300$ MeV at the 90\% CL). In this work we reply on $\rho_C$ and
$\phi_C$ to get a large complex $a_2$. } Therefore, there is no
reason to restrict $\rho_H$ to the range $0\leq \rho_H\leq 1$.

A sizable color-suppressed tree amplitude also can be induced via
color-allowed decay $B^-\to K^-\eta'$ followed by the rescattering
of $K^-\eta'$ into $K^-\pi^0$ as depicted in Fig. 1. Recall that
among the 2-body $B$ decays, $B\to K\eta'$ has the largest
branching fraction, of order $70\times 10^{-6}$. This final-state
rescattering has the same topology as the color-suppressed tree
diagram \cite{CCSfsi}. One of us (CKC) has studied the FSI effects
through residual rescattering among $PP$ states and resolved the
$B$-$CP$ puzzles \cite{Chua}.

\begin{figure}[t]
\begin{center}
\includegraphics[width=0.6\textwidth]{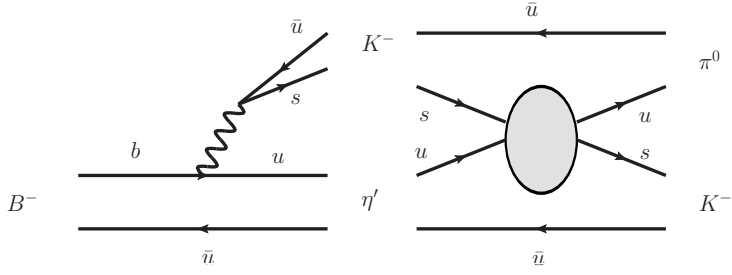}
\vspace{0.0cm}
\caption{Contribution to the color-suppressed tree amplitude of $B^-\to K^-\pi^0$ from the weak decay $B^-\to K^-\eta'$ followed by the final-state rescattering of $K^-\eta'$ into $\bar K^0\pi^0$. This has the same topology as the color-suppressed tree diagram.} \label{fig:trigluon} \end{center}
\end{figure}

 \begin{table}[tb]
 \caption{Same as Table \ref{tab:PP} except for some  selective $B\to VP$ decays with $\rho_C=0.8$ and $\phi_C=-80^\circ$.
 } \label{tab:VP}
\begin{ruledtabular}
 \begin{tabular}{l  c c c c c}
 {Modes}
   &   {\rm W/o}~$\rho_{A,C},\phi_{A,C}$ &  {\rm With}~$\rho_A,\phi_A$ &  {\rm With}~$\rho_{A,C},\phi_{A,C}$ & Expt. \cite{HFAG} \\  \hline
  $\B(\ov B^0\to K^-\rho^+)$ & $6.5^{+5.4+0.4}_{-2.6-0.4}$ & $8.6^{+5.7+7.4}_{-2.8-4.5}$ & $8.6^{+5.7+7.4}_{-2.8-4.5}$ & $8.6^{+0.9}_{-1.1}$ \\
  $\B(\ov B^0\to \bar K^0\rho^0)$ & $4.7^{+3.3+0.3}_{-1.7-0.3}$ & $5.5^{+3.5+4.3}_{-1.8-2.8}$ & $5.4^{+3.3+4.3}_{-1.7-2.8}$ & $5.4^{+0.9}_{-1.0}$ \\
  $\B(B^-\to \bar K^0\rho^-)$ & $5.5^{+6.1+0.7}_{-2.8-0.5}$ & $7.8^{+6.3+7.3}_{-2.9-4.4}$ & $7.8^{+6.3+7.3}_{-2.9-4.4}$ & $8.0^{+1.5}_{-1.4}$ \\
  $\B(B^-\to K^-\rho^0)$ & $1.9^{+2.5+0.3}_{-1.0-0.2}$ & $3.3^{+2.6+2.9}_{-1.1-1.7}$ & $3.5^{+2.9+2.9}_{-1.2-1.8}$ & $3.81^{+0.48}_{-0.46}$ \\
  $\B(\ov B^0\to K^{*-}\pi^+)$ & $3.7^{+0.5+0.4}_{-0.5-0.4}$ & $9.2^{+1.0+3.7}_{-1.0-3.3}$ & $9.2^{+1.0+3.7}_{-1.0-3.3}$  & $10.3\pm1.1$ \\
  $\B(\ov B^0\to \bar K^{*0}\pi^0)$ & $1.1^{+0.2+0.2}_{-0.2-0.2}$ & $3.5^{+0.4+1.7}_{-0.5-1.5}$ & $3.5^{+0.4+1.6}_{-0.4-1.4}$ & $2.4\pm0.7$ \\
  $\B(B^-\to \bar K^{*0}\pi^-)$ & $4.0^{+0.7+0.6}_{-0.9-0.6}$ & $10.4^{+1.3+4.3}_{-1.5-3.9}$ & $10.4^{+1.3+4.3}_{-1.5-3.9}$ & $9.9^{+0.8}_{-0.9}$ \\
  $\B(B^-\to K^{*-}\pi^0)$ & $3.2^{+0.4+0.3}_{-0.4-0.3}$ & $6.8^{+0.7+2.3}_{-0.7-2.2}$ & $6.7^{+0.7+2.4}_{-0.7-2.2}$ & $6.9\pm2.3$ \\
  $\B(\ov B^0\to \bar K^{*0}\eta)$ & $11.0^{+6.9+1.7}_{-3.5-1.0}$ & $15.4^{+7.7+9.4}_{-4.0-7.1}$& $15.6^{+7.9+9.4}_{-4.1-7.1}$ & $15.9\pm1.0$ \\
  $\B(\bar B^0\to \rho^0\pi^0)$ & $0.76^{+0.96+0.66}_{-0.37-0.31}$ & $0.58^{+0.88+0.60}_{-0.32-0.22}$ & $1.3^{+1.7+1.2}_{-0.6-0.6}$ & $2.0\pm0.5$ \footnotemark[1] \\
  $\B(B^-\to \rho^-\pi^0)$ & $11.6^{+1.2+0.9}_{-0.9-0.5}$ & $11.8^{+1.3+1.0}_{-0.9-0.6}$ & $11.8^{+1.8+1.4}_{-1.1-1.4}$ & $10.9^{+1.4}_{-1.5}$ \\
  $\B(B^-\to \rho^0\pi^-)$ & $8.2^{+1.8+1.2}_{-0.9-0.6}$ & $8.5^{+1.8+1.2}_{-0.9-0.6}$ & $8.7^{+2.7+1.7}_{-1.3-1.4}$ & $8.3^{+1.2}_{-1.3}$ \\
  $\B(\bar B^0\to \rho^-\pi^+)$ & $15.3^{+1.0+0.5}_{-1.5-0.9}$ & $15.9^{+1.1+0.9}_{-1.5-1.1}$ & $15.9^{+1.1+0.9}_{-1.5-1.1}$ & $15.7\pm1.8$ \\
  $\B(\bar B^0\to \rho^+\pi^-)$ & $8.4^{+0.4+0.3}_{-0.7-0.5}$ & $9.2^{+0.4+0.5}_{-0.7-0.7}$ & $9.2^{+0.4+0.5}_{-0.7-0.7}$ & $7.3\pm1.2$ \\
  \hline
  $\acp(\ov B^0\to K^-\rho^+)$ & $-1.3^{+0.7+3.8}_{-0.3-3.8}$ & $31.9^{+11.5+19.6}_{-11.0-12.7}$ & $31.9^{+11.5+19.6}_{-11.0-12.7}$ & $15\pm6$ \\
  $\acp(\ov B^0\to \bar K^0\rho^0)$ & $6.8^{+1.1+4.9}_{-1.2-4.9}$ & $-5.0^{+3.2+6.0}_{-6.4-4.5}$ & $8.7^{+1.2+8.7}_{-1.2-6.8}$ & $6\pm12$\\
  $\acp(B^-\to \bar K^0\rho^-)$ & $0.24^{+0.12+0.08}_{-0.15-0.07}$ & $0.27^{+0.19+0.46}_{-0.27-0.17}$ & $0.27^{+0.19+0.46}_{-0.27-0.17}$ & $-12\pm17$ \\
  $\acp(B^-\to K^-\rho^0)$ & $-8.3^{+3.5+7.0}_{-0.9-7.0}$ & $56.5^{+16.1+30.0}_{-18.2-22.8}$ & $45.4^{+17.8+31.4}_{-19.4-23.2}$  & $44^{+12}_{-17}$  \\
  $\acp(\ov B^0\to K^{*-}\pi^+)$ & $15.6^{+0.9+4.5}_{-0.7-4.7}$ & $-12.1^{+0.5+12.6}_{-0.5-16.0}$ & $-12.1^{+0.5+12.6}_{-0.5-16.0}$ & $-23\pm8$ \\
  $\acp(\ov B^0\to \bar K^{*0}\pi^0)$ & $-12.0^{+2.4+11.3}_{-4.6-~7.6}$ & $-0.87^{+1.71+6.04}_{-0.89-6.79}$ & $-10.7^{+1.8+9.1}_{-2.8-6.3}$  & $-15\pm12$ \\
  $\acp(B^-\to \bar K^{*0}\pi^-)$ & $0.97^{+0.11+0.12}_{-0.07-0.11}$ & $0.39^{+0.04+0.10}_{-0.03-0.12}$ & $0.39^{+0.04+0.10}_{-0.03-0.12}$ & $3.2\pm5.4$ \\
  $\acp(B^-\to K^{*-}\pi^0)$ & $17.5^{+2.0+6.3}_{-1.3-8.0}$ & $-6.7^{+0.7+11.8}_{-1.1-14.0}$ & $1.6^{+3.1+11.1}_{-1.7-14.3}$ & $4\pm29$ \\
  $\acp(\ov B^0\to \bar K^{*0}\eta)$ & $3.0^{+0.4+1.9}_{-0.4-1.8}$ & $0.20^{+0.51+2.00}_{-1.00-1.21}$ & $3.5^{+0.4+2.7}_{-0.5-2.4}$ & $19\pm5$ \\
  $\acp(\bar B^0\to \rho^0\pi^0)$ & $-2.3^{+2.4+9.9}_{-3.7-9.2}$ & $31.5^{+13.3+21.5}_{-12.5-30.9}$ & $11.0^{+5.0+23.5}_{-5.7-28.8}$ & $-30\pm38$ \footnotemark[2] \\
  $\acp(B^-\to \rho^-\pi^0)$ & $-5.4^{+0.4+2.0}_{-0.3-2.1}$ & $16.3^{+1.1+~7.1}_{-1.2-10.5}$ & $9.7^{+2.1+~8.0}_{-3.1-10.3}$ & $2\pm11$ \\
  $\acp(B^-\to \rho^0\pi^-)$ & $6.7^{+0.5+3.5}_{-0.8-3.1}$ & $-19.8^{+1.7+12.6}_{-1.2-8.8}$ & $-9.8^{+3.4+11.4}_{-2.6-10.2}$ & $18^{+~9}_{-17}$ \\
  $\acp(\bar B^0\to \rho^-\pi^+)$ & $-3.5^{+0.2+1.0}_{-0.2-0.9}$ & $4.4^{+0.3+5.8}_{-0.3-6.8}$ & $4.4^{+0.3+5.8}_{-0.3-6.8}$ & $11\pm6$ \\
  $\acp(\bar B^0\to \rho^+\pi^-)$ & $0.6^{+0.1+2.2}_{-0.1-2.2}$ & $-22.7^{+0.9+8.2}_{-1.1-4.4}$ & $-22.7^{+0.9+8.2}_{-1.1-4.4}$ & $-18\pm12$ \\
 \end{tabular}
 \footnotetext[1]{If an $S$ factor is included, the average will become $2.0\pm0.8$.}
\footnotetext[2]{This is the average of $10\pm40\pm53$ by BaBar \cite{BaBarrho0pi0acp} and $-49\pm36\pm28$ by Belle \cite{Bellerho0pi0acp}.}
 \end{ruledtabular}
 \end{table}

\vskip 0.4cm {\bf 4.}~Power corrections to $a_2$ for $B\to VP$ and
$B\to VV$ are not the same as that for $B\to PP$ as described by
Eq. (\ref{eq:a2PP}). From Table \ref{tab:VP} we see that an
enhancement of $a_2$ is needed to improve the rates of $B\to
\rho^0\pi^0$ and the direct \CP asymmetry of $\bar B^0\to \bar
K^{*0}\eta$. However, it is constrained by the measured rates of
$\rho^0\pi^-$ and $\rho^-\pi^0$ modes. This means that
$\rho_C(VP)$ is preferred to be smaller than $\rho_C(PP)=1.3$\,.
In Table \ref{tab:VP} we show the branching fractions and \CP
asymmetries in $B\to VP$ decays for $\rho_C(VP)=0.8$\, and $\phi_C(VA)=-80^\circ$. The
corresponding values of $a_2(VP)$ are
\be \label{eq:a2VP}
&& a_2(\pi\rho)\approx 0.40\,e^{-i51^\circ}, \quad a_2(\rho\pi)\approx 0.38\,e^{-i52^\circ}, \non \\
 && a_2(\rho \bar K)\approx 0.36\,e^{-i52^\circ}, \quad
a_2(\pi \bar K^*)\approx 0.39\,e^{-i51^\circ}.
\en

It is clear from Table \ref{tab:VP} that in the heavy quark limit,
the predicted rates for $\bar B\to \bar K^*\pi$ are too small by a
factor of $2\sim 3$, while $\B(\bar B\to \bar K\rho)$ are too
small by $(15\sim 50)\%$ compared with experiment. The rate
deficit for penguin-dominated decays can be accounted by the
subleading power corrections from penguin annihilation. Soft
corrections to $a_2$ will enhance $\B(B\to \rho^0\pi^0)$ to the
order of $1.3\times 10^{-6}$, while the BaBar and Belle results,
$(1.4\pm0.6\pm0.3)\times 10^{-6}$ \cite{BaBarrho0pi0} and
$(3.0\pm0.5\pm0.7)\times 10^{-6}$ \cite{Bellerho0pi0}
respectively, differ in their central values by a factor of 2.
Improved measurements are certainly needed for this decay mode. As
for direct \CP asymmetries, we see that penguin annihilation will
flip the sign of $\acp(K^-\rho^0)$ into the right direction. Power
corrections to the color-suppressed tree amplitude are needed to
improve the prediction for $\acp(\bar K^{*0}\eta)$. Our prediction
is of order 0.035 to be compared with the experimental value of
$0.19\pm0.05$. The pQCD prediction of $\acp(\bar K^{*0}\eta)\sim
0.0057$ \cite{XiaoVeta} is too small, while the SECT result of
$\sim-0.01$ \cite{SCETVP} has a wrong sign. For $\acp(\bar
K^0\rho^0)$, it gets a sign flip after including soft effects on
$a_2$. Our prediction is $(8.7^{+8.8}_{-6.9})\%$, while it is
$0.06\pm0.20$ experimentally. Defining $\Delta A_{K^*\pi}\equiv
\acp(K^{*-}\pi^0)-\acp(K^{*-}\pi^+)$ in analog to $\Delta
A_{K\pi}$, we predict that $\Delta
A_{K^*\pi}=(13.7^{+2.9+3.6}_{-1.4-6.9})\%$, while it is naively
expected that $K^{*-}\pi^0$ and $K^{*-}\pi^+$ have similar {\it CP}-violating effects.
It is of importance to measure \CP asymmetries of these
two modes to test our prediction. For mixing-induced \CP
violation, we obtain $\Delta S_{\phi K_S}=
0.022^{+0.004}_{-0.002}$\,, $\Delta S_{\omega K_S}=
0.17^{+0.06}_{-0.08}$\, and $\Delta S_{\rho^0K_S}=
-0.17^{+0.09}_{-0.18}$  \cite{CCSnew}, where $\Delta
S_f\equiv-\eta_f S_f-\sin 2 \beta$. It turns out that soft
corrections to $a_2$ have significant effects on the last two
quantities.

As for $B\to VV$ decays, we notice that the calculated  $B^0\to
\rho^0\rho^0$ rate in QCDF  is $\B(B^0\to
\rho^0\rho^0)=(0.88^{+1.46+1.06}_{-0.41-0.20})\times 10^{-6}$ for
$\rho_C=0$  \cite{ChengVV},  while the world average is
$(0.73^{+0.27}_{-0.28})\times 10^{-6}$ \cite{HFAG}. Therefore,
soft power correction to $a_2$ or $\rho_C(VV)$ should be small for
$B^0\to\rho^0\rho^0$. Consequently, a pattern follows: Effects of
power corrections on $a_2$ are large for $PP$ modes, moderate for
$VP$ ones and very small for $VV$ cases. \footnote{Since the
chiral factor $r_\chi^V$ for the vector meson is substantially
smaller than $r_\chi^P$ for the pseudoscalar meson (typically,
$r_\chi^P={\cal O}(0.8)$ and $r_\chi^V={\cal O}(0.2)$ at the hard collinear scale $\mu=\sqrt{\Lambda m_b}$), one may
argue that Eq. (\ref{eq:hardspec}) naturally explains  why the
power corrections to $a_2$ is smaller when $M_1$ is a vector
meson, provided that soft corrections arise from spectator
rescattering. Unfortunately, this is not the case. Numerically, we
found that, for example, $H(K^*\pi)$ is comparable to $H(K\pi)$.
This is due to the fact that $\int_0^1 dx \,r_\chi^M\Phi_m(x)/\bar
x$ is equal to $X_Hr_\chi^P$ for $M=P$ and approximated to
$3(X_H-2)r_\chi^V$ for $M=V$. } This is consistent with the
observation made in \cite{Kagan} that soft power correction
dominance is much larger for $PP$ than $VP$ and $VV$ final states.
It has been argued that this has to do with the special nature of
the pion which is a $q\bar q$ bound state on the one hand and a
nearly massless Nambu-Goldstone boson on the other hand
\cite{Kagan}. The two seemingly distinct pictures of the pion can
be reconciled by considering a soft cloud of higher Fock states
surrounding the bound valence quarks.  From the FSI point of view,
since $B\to\rho^+\rho^-$ has a rate much larger than $B\to
\pi^+\pi^-$, it is natural to expect that $B\to\pi^0\pi^0$
receives a large enhancement from the weak decay
$B\to\rho^+\rho^-$ followed by the rescattering of $\rho^+\rho^-$
to $\pi^0\pi^0$ through the exchange of the $\rho$ particle.
Likewise, it is anticipated that $B\to \rho^0\rho^0$ will receive
a large enhancement via isospin final-state interactions from
$B\to \rho^+\rho^-$. The fact that the branching fraction of this
mode is rather small and is consistent with the theory prediction
implies that the isospin phase difference of $\delta_0^\rho$ and
$\delta_2^\rho$ and the final-state interaction must be negligible
 \cite{Vysotsky}.

\vskip 0.4cm {\bf 5.}~$B$-{\it CP} puzzles arise in the framework
of QCD factorization because power corrections due to penguin
annihilation, that account for the observed  rates of
penguin-dominated two-body decays of $B$ mesons and direct \CP
asymmetries $\acp(K^-\pi^+)$, $\acp(K^{*-}\pi^+)$,
$\acp(K^-\rho^0)$ and $\acp(\pi^+\pi^-)$, will flip the signs of
direct {\it CP}-violating effects in  $B^-\to K^-\pi^0, B^-\to
K^-\eta, B^-\to \pi^-\eta$ and $\bar B^0\to\pi^0\pi^0$ to wrong
ones when confronted with experiment. We have shown that power
corrections to the color-suppressed tree amplitude due to hard
spectator interactions and/or final-state interactions will yield
correct signs again for  aforementioned \CP asymmetries  and
accommodate the observed $\pi^0\pi^0$ and $\rho^0\pi^0$ rates
simultaneously. {\it CP}-violating asymmetries of $B^-\to K^-\eta$
can be understood as a consequence of soft corrections to $a_2$.
$\acp(\bar K^0\pi^0)$ is predicted to be of order $-0.10$\,, in
agreement with that inferred from the {\it CP}-asymmetry sum rule,
or SU(3) relation or the diagrammatical approach. For direct \CP
violation in $B^-\to K^{*-}\eta, \pi^-\eta$, our predictions are
in better agreement with experiment than pQCD and SCET. For $\bar
B^0\to\bar K^0\rho^0$, we obtained $\acp(\bar
K^0\rho^0)=0.087^{+0.088}_{-0.069}$. We argued that the smallness
of \CP asymmetry of $B^-\to \pi^-\pi^0$ is not affected by the
soft corrections under consideration. For the  \CP asymmetry
difference in $K^*\pi$ modes defined by $\Delta A_{K^*\pi}\equiv
\acp(K^{*-}\pi^0)-\acp(K^{*-}\pi^+)$, we predict that $\Delta
A_{K^*\pi}\sim 14\%$, while these two modes are naively expected
to have similar direct {\it CP}-violating effects. For
mixing-induced \CP violation, we found $\Delta S_{\pi^0K_S}=
0.12^{+0.07}_{-0.06}$\,, $\Delta S_{\phi K_S}=
0.022^{+0.004}_{-0.002}$\,, $\Delta S_{\omega K_S}=
0.17^{+0.06}_{-0.08}$\, and $\Delta S_{\rho^0K_S}=
-0.17^{+0.09}_{-0.18}$.

\vskip 1.60cm {\bf Acknowledgments}

We are grateful to Chuan-Hung Chen, Cheng-Wei Chiang, Hsiang-nan
Li, Tri-Nang Pham and Amarjit Soni for useful discussions. One of
us (H.Y.C.) wishes to thank the hospitality of the Physics
Department, Brookhaven National Laboratory. This research was
supported in part by the National Science Council of R.O.C. under
Grant Nos. NSC97-2112-M-001-004-MY3 and NSC97-2112-M-033-002-MY3.



\begin{thebibliography}{99}

\bibitem{ChengSmith} H.Y. Cheng and J. Smith, Annu. Rev. Nucl.  Part. Sci. {\bf 59}, 215 (2009) [arXiv:0901.4396 [hep-ph]].



\bibitem{BBNS} M. Beneke, G. Buchalla, M. Neubert, and C.T. Sachrajda,
\prl {\bf 83}, 1914 (1999); \np B {\bf 591}, 313 (2000).

\bibitem{pQCD}
Y.Y. Keum, H.-n. Li, and A.I. Sanda,  \pr D { \bf 63}, 054008 (2001).

\bibitem{BN} M. Beneke and M. Neubert, \np B {\bf 675}, 333
(2003).

\bibitem{PDG} Particle Data Group, C. Amsler {\it et al}., Phys. Lett. B {\bf 667}, 1 (2008).

\bibitem{HFAG}
E. Barberio et al. (Heavy Flavor Averaging Group),
 arXiv:0704.3575 [hep-ex] (2007) and online update at
 http://www.slac.stanford.edu/xorg/hfag.

\bibitem{BaBar:Beta}
  B.~Aubert  {\it et al.} [BABAR Collaboration],
  arXiv:0907.1743 [hep-ex].

\bibitem{Li05}
Y.~Y.~Charng and H.~n.~Li,
  Phys.\ Rev.\  D {\bf 71}, 014036 (2005);
 H.-n. Li, S. Mishima, and A.I. Sanda,
\pr D {\bf 72}, 114005 (2005).

\bibitem{Kim}
  C.~S.~Kim, S.~Oh and C.~Yu,
  Phys.\ Rev.\  D {\bf 72}, 074005 (2005)
  [arXiv:hep-ph/0505060].

\bibitem{Rosner}
  M.~Gronau and J.~L.~Rosner,
  Phys.\ Lett.\  B {\bf 644}, 237 (2007)
  [arXiv:hep-ph/0610227].

\bibitem{Chua}
  C.~K.~Chua,
  Phys.\ Rev.\  D {\bf 78}, 076002 (2008)
  [arXiv:0712.4187].

\bibitem{Ciuchini}
  M.~Ciuchini, E.~Franco, G.~Martinelli, M.~Pierini and L.~Silvestrini,
  Phys.\ Lett.\  B {\bf 674}, 197 (2009).

\bibitem{Kagan}
  M.~Duraisamy and A.~L.~Kagan,
  arXiv:0812.3162 [hep-ph].

 \bibitem{Li09}
  H.~n.~Li and S.~Mishima,
  arXiv:0901.1272 [hep-ph].

\bibitem{Baek09}
 S.~Baek, C.~W.~Chiang, M.~Gronau, D.~London and J.~L.~Rosner,
  arXiv:0905.1495 [hep-ph].

\bibitem{LargeEWP} T. Yoshikawa,
\pr D {\bf 68}, 054023 (2003); S.~Mishima and T.~Yoshikawa,
  Phys.\ Rev.\  D {\bf 70}, 094024 (2004);
A.J. Buras, R. Fleischer, S. Recksiegel, F. Schwab,
\prl {\bf 92}, 101804 (2004); Y.L. Wu and Y.F. Zhou,
\pr D {\bf 72}, 034037 (2005); S. Baek,
 P.~Hamel, D.~London, A.~Datta and D.~A.~Suprun,
\pr D {\bf 71}, 057502 (2005);   S. Baek and  D. London,
\pl B {\bf 653}, 249 (2007); T. Feldmann, M. Jung, and T. Mannel,
JHEP {\bf 0808}, 066 (2008).

  \bibitem{Hou}
  W.~S.~Hou, H.~n.~Li, S.~Mishima and M.~Nagashima,
  Phys.\ Rev.\ Lett.\  {\bf 98}, 131801 (2007)
  [arXiv:hep-ph/0611107];
  A.~Soni, A.~K.~Alok, A.~Giri, R.~Mohanta and S.~Nandi,
  arXiv:0807.1971 [hep-ph].

\bibitem{Chiang}
  C.~W.~Chiang, M.~Gronau, J.~L.~Rosner and D.~A.~Suprun,
  Phys.\ Rev.\  D {\bf 70}, 034020 (2004)
  [arXiv:hep-ph/0404073]; C.~W.~Chiang and Y.~F.~Zhou,
  JHEP {\bf 0612}, 027 (2006)
  [arXiv:hep-ph/0609128].


\bibitem{Beneke}
  M.~Beneke and S.~Jager,
  Nucl.\ Phys.\  B {\bf 751}, 160 (2006)
  [arXiv:hep-ph/0512351];  N.~Kivel,
  JHEP {\bf 0705}, 019 (2007)
  [arXiv:hep-ph/0608291];
 V.~Pilipp,
  Nucl.\ Phys.\  B {\bf 794}, 154 (2008).

\bibitem{Bell}
  G.~Bell,
  Nucl.\ Phys.\  B {\bf 795}, 1 (2008)
  [arXiv:0705.3127 [hep-ph]]; 
  arXiv:0902.1915 [hep-ph].

\bibitem{BenekeFPCP} M. Beneke, talk presented at the FPCP2008 Conference on Flavor Physics and CP violation, May 5-9, 2008, Taipei, Taiwan.

\bibitem{CCSnew} H.Y.~Cheng and C.K.~Chua, in preparation.



\bibitem{Lunghi}
  E.~Lunghi and A.~Soni,
  JHEP {\bf 0709}, 053 (2007)
  [arXiv:0707.0212 [hep-ph]].

\bibitem{BaBarK0pi0}
  B.~Aubert {\it et al.}  [BABAR Collaboration],
  Phys.\ Rev.\  D {\bf 79}, 052003 (2009)
  [arXiv:0809.1174 [hep-ex]].

\bibitem{BelleK0pi0}
  I.~Adachi {\it et al.}  [Belle Collaboration],
  arXiv:0809.4366 [hep-ex].

\bibitem{Deshpande} N.G. Deshpande and X.G. He, \prl {\bf 75}, 1703
(1995).

\bibitem{AS98} D. Atwood and A. Soni, \pr D {\bf 58}, 036005 (1998);
  M.~Gronau,
  Phys.\ Lett.\  B {\bf 627}, 82 (2005).

\bibitem{Chiang09}
  S.~Baek, C.~W.~Chiang and D.~London,
  arXiv:0903.3086 [hep-ph].


\bibitem{CCSsin2beta} H.Y.~Cheng, C.K.~Chua, and A.~Soni, Phys. Rev. D {\bf
72}, 014006 (2005).

\bibitem{Fleischer}
  R.~Fleischer, S.~Jager, D.~Pirjol and J.~Zupan,
  Phys.\ Rev.\  D {\bf 78}, 111501 (2008)
  [arXiv:0806.2900 [hep-ph]]; M.~Gronau and J.~L.~Rosner,
  Phys.\ Lett.\  B {\bf 666}, 467 (2008)
  [arXiv:0807.3080 [hep-ph]].

\bibitem{BaBarpi0pi0}
  B.~Aubert {\it et al.}  [BABAR Collaboration],
  arXiv:0807.4226 [hep-ex].

\bibitem{Bellepi0pi0}
  K.~Abe {\it et al.}  [Belle Collaboration],
  arXiv:hep-ex/0610065.

\bibitem{CCSfsi} H.Y.~Cheng, C.K.~Chua, and A.~Soni, Phys. Rev. D {\bf
71}, 014030 (2005).


\bibitem{BSS} M. Bander, D. Silverman, and A. Soni, \prl {\bf 43} 242 (1979);
S. Barshay and G. Kreyerhoff, \pl B {\bf 578}, 330 (2004).


\bibitem{fetac}
  A.~Ali, J.~Chay, C.~Greub and P.~Ko,
  Phys.\ Lett.\  B {\bf 424}, 161 (1998)
  [arXiv:hep-ph/9712372]; M.~Franz, M.~V.~Polyakov and K.~Goeke,
  Phys.\ Rev.\  D {\bf 62}, 074024 (2000)
  [arXiv:hep-ph/0002240];  M.~Beneke and M.~Neubert,
  Nucl.\ Phys.\  B {\bf 651}, 225 (2003)
  [arXiv:hep-ph/0210085].

\bibitem{FKS} T. Feldmann, P. Kroll, and B. Stech, Phys. Rev. D
{\bf 58}, 114006 (1998); Phys. Lett. B {\bf 449}, 339 (1999).

\bibitem{ChenKeta} A.G. Akeroyd, C.H. Chen, and C.Q. Geng,  \pr D {\bf 75} 054003 (2007).

\bibitem{Xiao} Z.~J.~Xiao, Z.~Q.~Zhang, X.~Liu and L.~B.~Guo,
  Phys.\ Rev.\  D {\bf 78}, 114001 (2008).

\bibitem{Xiaopieta}
  H.~S.~Wang, X.~Liu, Z.~J.~Xiao, L.~B.~Guo and C.~D.~Lu,
  Nucl.\ Phys.\  B {\bf 738}, 243 (2006).

\bibitem{Zupan}
  A.~R.~Williamson and J.~Zupan,
  Phys.\ Rev.\  D {\bf 74}, 014003 (2006)
  [Erratum-ibid.\  D {\bf 74}, 03901 (2006)].

\bibitem{BBNS2} M. Beneke, G. Buchalla, M. Neubert, and C.T. Sachrajda,
\np B {\bf 606}, 245 (2001).

\bibitem{Bell09}
  G.~Bell and V.~Pilipp,
  arXiv:0907.1016 [hep-ph].

\bibitem{BaBar:gammalnu}
  B.~Aubert {\it et al.}  [BABAR Collaboration],
  arXiv:0907.1681 [hep-ex].




\bibitem{BaBarrho0pi0acp}
  B.~Aubert {\it et al.}  [BABAR Collaboration],
  Phys.\ Rev.\  D {\bf 76}, 012004 (2007).

\bibitem{Bellerho0pi0acp}
  A.~Kusaka {\it et al.}  [Belle Collaboration],
  Phys.\ Rev.\ Lett.\  {\bf 98}, 221602 (2007).


\bibitem{BaBarrho0pi0}
  B.~Aubert {\it et al.}  [BABAR Collaboration],
  Phys.\ Rev.\ Lett.\  {\bf 93}, 051802 (2004)
  [arXiv:hep-ex/0311049].

\bibitem{Bellerho0pi0}
  A.~Kusaka {\it et al.}  [Belle Collaboration],
  Phys.\ Rev.\  D {\bf 77}, 072001 (2008)
  [arXiv:0710.4974 [hep-ex]].

\bibitem{XiaoVeta}
  Z.~Q.~Zhang and Z.~J.~Xiao,
  arXiv:0807.2024 [hep-ph].

\bibitem{SCETVP}
  W.~Wang, Y.~M.~Wang, D.~S.~Yang and C.~D.~Lu,
  Phys.\ Rev.\  D {\bf 78}, 034011 (2008).


\bibitem{ChengVV} H.Y. Cheng and K.C. Yang, \pr D {\bf 78}, 094001 (2008).





\bibitem{Vysotsky} A.B. Kaidalov and M.I. Vysotsky, \pl B {\bf 652}, 203 (2007).




\end{thebibliography}
\end{document}